\documentclass[twocolumn]{revtex4}
\usepackage{graphicx}
\usepackage[]{epsfig}

\newcommand{\beq}{\begin{equation}}
\newcommand{\eeq}{\end{equation}}
\newcommand{\beqd}{\begin{displaymath}}
\newcommand{\eeqd}{\end{displaymath}}
\newcommand{\beqa}{\begin{eqnarray}}
\newcommand{\eeqa}{\end{eqnarray}}
\newcommand{\non}{\nonumber}
\renewcommand{\a}{\alpha}
\renewcommand{\b}{\beta}
\newcommand{\D}{\Delta}
\newcommand{\e}{\epsilon}
\newcommand{\sign}{{\rm sign}}

\newcommand{\arctanh}{{\rm arctanh}}

\begin{document}
\title{Zero-Temperature Limit of the SUSY-breaking Complexity 
\\
in Diluted Spin-Glass Models}


\author{Giorgio Parisi$^{1}$ and Tommaso Rizzo$^{2}$}

\affiliation{$^{1}$Dipartimento di Fisica, Universit\`a di Roma ``La Sapienza'', INFN  and
Center for Statistical Mechanics and Complexity, INFM Roma 1 
P.le Aldo Moro 2, 00185 Roma,  Italy
\\
$^{2}$Laboratoire de Physique Th\'{e}orique de l'Ecole
Normale Sup\'{e}rieure, 24 rue Lhomond, 75231 Paris, France}

\begin{abstract}
We study the SUSY-breaking complexity of the Bethe Lattice Spin-Glass in the zero temperature limit.
We consider both the Gaussian and the bimodal distribution of the coupling constants.  For
$J_{ij}=\pm 1$ the SUSY breaking theory yields fields distributions that concentrate on integer
values at low temperatures, at variance with the unbroken SUSY theory.  This concentration takes
place both in the quenched as well as in the simpler annealed formulation.
\end{abstract}

\maketitle

The complexity, {\it i.e.} the logarithm of the number of metastable states, is a crucial tool to
understand spin-glasses and related disordered systems \cite{MPV}.  Recently it become clear that in
mean-field models there are two possibilities: a supersymmetric (SUSY) theory and a spontaneously
SUSY-broken theory.  The behavior of $\Sigma(f)$, the complexity of the states with free energy
$f$, in one-step replica-symmetry-breaking (1RSB) models is the following : at low free energies
(including the equilibrium one) the theory is supersymmetric (SUSY) while at higher free energies
the SUSY-breaking theory must be used \cite{isp} .  In models with full-RSB (FRSB) like the
Sherrington-Kirkpatrick model the SUSY solution is unstable immediately above the equilibrium free
energy $f_{eq}$ \cite{noiquen} and the complexity must be computed using a SUSY-breaking approach
at all $f>f_{eq}$.  The paradigm of the SUSY-breaking solution is the Bray and Moore computation of
the SK model that in the replica formulation corresponds to the two-group ansatz \cite{BMan}.  Some
apparent inconsistencies of this theory \cite{var} have been solved by the discovery of a vanishing
isolated eigenvalue \cite{ABM} in the spectrum of the TAP Hessian and in \cite{PR} it has been
proved that this result is a consequence of SUSY breaking and therefore is very general.  Later
these results have also received some numerical confirmations \cite{CGPprl}.  Thus the relevant
states are marginal within the SUSY-breaking theory but there are also non-marginal states with a
lower total complexity \cite{R1}.  The theory of diluted spin-glass models and optimization
problems has received a considerable boost by the application of the cavity method \cite{MP1,MPZ}.
The standard computation scheme of the complexity used in this context \cite{Moncomp} corresponds to
the SUSY theory and becomes unstable in certain regions of the parameter space \cite{MRpMRP}.  Much
as in mean-field models with infinite connectivity, this instability could be caused either by a
1RSB/FRSB transition or by a less standard SUSY/SUSY-breaking transition.  This has motivated the
search for a formulation of the SUSY-breaking theory within the cavity method which has been
achieved in \cite{R1}, see also \cite{CGPcav}.  If we add a small perturbation to the weight of the
states \cite{R1}, we select a set of non-marginal states slightly different from the relevant ones,
and the cavity method can be applied.  The relevant states are recovered sending to zero the
perturbation; this limit is non-trivial, indeed as the perturbation goes to zero the reweighing
associated to its presence remains finite because of marginality.  Later this method has been used
to derive the SUSY breaking theory for diluted models \cite{R2}.  Here we study the zero
temperature limit of this theory which is particularly interesting for the application to
optimization models.  Particularly in the case of bimodal distribution of the coupling constants
this limit is rather non-trivial, and our main result is that the field distributions concentrate
over integer values.  Remarkably this happens already within the so-called factorized approximation,
a result not totally unexpected because it has been recognized that it corresponds to the annealed
solution.

{\it The Complexity at Finite Temperature} - In the following we recall the theory at finite
temperature \cite{R2}.  One finds the recursive equations for the fields $h$ and $z$ in the Bethe
lattice spin-glass are: 
\beq h_0=\sum_{i=1}^{k}u(J_i,h_i)\ \ ; \ \
z_0=\sum_{i=1}^{k}{du(J_i,h_i)\over dh_i}z_i \,
\label{z0fin}
\eeq
where 
$u(J,h)\equiv \beta^{-1}\arctanh [\tanh (\beta J)\tanh (\beta h)]$.

The complexity is the number of the solutions of these equations.  The fields $z_i$ are zero in the
standard SUSY solution, they are non-zero only if the solutions are marginally stable (it can be
argued that they are proportional to the components of the eigenvector of the vanishing isolated
eigenvalue).

The recursive equation for the distribution of the fields $h_0$ and $z_0$ is 
\beqd
P_0(h_0,z_0) = 
C \int \prod_{i=1}^{k}P_i(h_i,z_i)e^{u \beta \Delta F^{(iter)}+ \Delta X^{(iter)}} \times
\eeqd
\beq
 \times  \delta \left(h_0-\sum_{i=1}^{k}u(J_i,h_i)\right)\delta\left(z_0-\sum_{i=1}^{k}{du(J_i,h_i)\over dh_i}
 z_i\right)dh_idz_i, 
\label{Ph0z0}
\eeq
where:
\beq
-\beta \D F^{(iter)}=\ln \left[2 \cosh \beta h_0 \right]+\sum_{i=1}^{k}\ln \left[  {\cosh \beta J_i  \over \cosh  \beta u(J_i,h_i)} \right]
\label{df1}
\eeq
\beq
\Delta X^{(iter)}=\sum_{i=1}^{k}\left[ \tanh \beta h_0 -\tanh \beta J_i \tanh \beta h_i       \right]{d u \over d h_i} z_i
\label{dx11}
\eeq
The shifts in the process of merging $k+1$ lines, $\Delta F^{(1)}$ and $\Delta X^{(1)}$, are the
same as above with $k$ replaced by $k+1$.
The distribution of distributions ${\mathcal P}[P(h,z)]$ can be determined numerically by population
dynamics \cite{MP1}; as usual in the
factorized approximation  the population $P_i(h_i,z_i)$ is site independent and corresponds to
the annealed theory \cite{R2}.
The functional $\Phi (u)=\Sigma (f)- u \beta f$ is the Legendre transform of the complexity; its shift is given by:
\beq
\Delta \Phi(u)^{(iter)}=\ln \int \prod_{i=1}^{k}P_i(h_i,z_i)dh_idz_i\ e^{u \beta \Delta F^{(iter)}+\Delta X^{(iter)}}
\label{dS1b}
\eeq
The Legendre transform with respect to $u$ of the complexity is given by
\beqa
\Phi(u)&=&{k+1\over 2}\int \prod_{i=1}^{k}{\mathcal {P}}(P_i(h_i,z_i))\langle \Delta \Phi(u)^{(iter)}\rangle_J
\non
\\
&-&{k-1\over 2}\int \prod_{i=1}^{k+1}{\mathcal {P}}(P_i(h_i,z_i))\langle \Delta \Phi(u)^{(1)}\rangle_J
\label{comp1}
\eeqa
where the square brackets mean average over the bonds $J_i$, and $\Delta \Phi^{(1)}$ has the same expression  $\Delta \Phi^{(iter)}$
with $k+1$ indexes instead of $k$.
At $u=0$ eq. (\ref{comp1}) gives the total complexity.
The free energy is given by $\beta f= d\Phi /du$.  This derivative can be computed numerically or
more efficiently by deriving with respect to $u$ the variational expression of $\Phi(u)$, eq.  35 in
\cite{R2}.

{\it The Zero-Temperature Limit}
- {\it Gaussian Distribution of the $J_{ij}$} -
In the $T\rightarrow 0$ limit we have to modify the functions $u(h_i,J_i)$ and $du/dh$ entering the recursion equations for the fields, 
\beqa
\lim_{T\rightarrow 0} u(J,h)&=& \sign \, J_i  \left\{
\begin{array}{ccc}
h_i &  {\rm if} &  |h_i|<|J_i|
\\
|J_i| \sign h_i &  {\rm if} &  |h_i|>|J_i|
\end{array}
\right. 
\\
\lim_{T\rightarrow 0} {du \over dh}&=& \sign \, J_i  \left\{
\begin{array}{ccc}
1 &  {\rm if} &  |h_i|<|J_i|
\\
0  &  {\rm if} &  |h_i|>|J_i|
\end{array}
\right. 
\eeqa
The limit of the shifts are obtained simply replacing the functions $\tanh \beta h_i$ with $\sign\,
h_i$ in the expressions (\ref{df1}) and (\ref{dx11}), thus we obtain
\beqa
- \D F^{(1)}&=&|h_0|+\sum_{i=1}^{k+1}(|J_i|-|u(J_i,h_i)| )\\
\Delta X^{(1)}&=&\sum_{i=1}^{k+1}\left[ \sign h_0 -\sign (J_i  h_i)       \right]{d u \over d h_i} z_i
\eeqa
Using this expression we can evaluate the distributions of the fields at zero temperature through population dynamics.
In general the resulting distributions of the fields have the following properties: i) there is a finite probability of having a strictly zero value of $z$, ii) the non-zero values of $z$ are rather large and diverging with the connectivity, iii) there is a strong correlation between $z_i$ and $h_i$ such that $\sign (h_i z_i)$ has a very small probability of being negative.

{\it Bimodal Distribution} ($J_{ij}=\pm 1$). 
In this case the fields concentrate over the integers in the $T\rightarrow 0$ limit but we must keep track of the behavior of the finite-temperature corrections to the integer value.
We rewrite the field $h_i$ as the sum of an integer contribution
$k_i$ and an infinitesimal contribution $T \epsilon_i$:
$h_i=k_i+ T \epsilon_i+ O(T^2)$.
The fields $\e_i$  remain finite in the $T\rightarrow 0$ limit.
We  must consider populations of  triplets $\{k_i,\e_i,z_i\}$ obeying the following recursion equations:
\beqa
k_0&=&\sum_{i=1}^{k} J_i \sign(k_i)
\label{reck}
\\
\epsilon_0&=&\sum_{i=1}^{k} \sign(J_i)g (k_i,\epsilon_i)
\label{rece}
\\
z_0&=&\sum_{i}^{k} {du \over dh_i}z_i=
\sum_{i}^{k}\left| {du \over dh_i}\right| \sign(J_i)z_i
\label{recz}
\eeqa
where the function $g(k_i,\epsilon_i)$ can be obtained taking the limit of the expression of $u(J_i,h_i)$ and is given by:
\beq
g(k_i,\e_i)= 
\left\{
\begin{array}{ccc}
\e_i  & {\rm for} & k_i=0
\\
{\e_i \over 2}-{\sign(k_i)\over 2}\ln 2 \cosh \e_i & {\rm for} & |k_i|=1
\\
0 & {\rm for} & |k_i|>1 
\end{array}
\right.
\label{ge}
\eeq
\beq
\left| {du \over dh_i}\right| =
\left\{
\begin{array}{ccc}
1  & {\rm for} & k_i=0
\\
{1 \over 2}-{\sign(k_i)\over 2}\tanh  \e_i & {\rm for} & |k_i|=1
\\
e^{-2 \b (|h_i|-|J_i|)} \simeq  0 & {\rm for} & |k_i|>1 
\end{array}
\right.
\label{dudq}
\eeq The $\e_i$'s have a precise meaning at zero temperature: if the spin $0$ is a {\it spin-fous},
{\it i.e.} $k_0=0$ the quantity $\tanh \e_0$ is its magnetization averaged over the many
configurations of the same energy that represent the zero temperature state.  At low temperatures
the behavior of the $z$ fields is dramatically different depending on the parity of $k$.  If $k$ is
even, the $z$ fields remains finite in the $T\rightarrow 0$ limit (they would diverge in the
high-connectivity limit).  Instead if $k$ is odd, the $z$-fields diverge as $\exp(2\beta|J|)$ and must
be rescaled at zero temperature.

\begin{figure}[htb]
\begin{center}
\epsfig{file=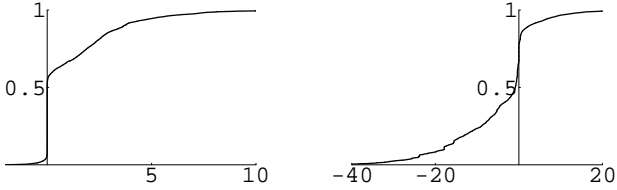,width=9cm} 
\caption{$T\rightarrow 0$ limit annealed theory for Bimodal Distribution, k=2. Left: cumulative distribution function of $\sign (k_i) z_i$, Right: cumulative distribution function of the $\e_i$ at $k_i=1$}
\label{figure1}
\end{center}\end{figure}

{\it Even Values of $k$} - 
In this case the $z$-fields remain finite. 
In order to compute the reweighing term eq. (\ref{dx11}) in the zero temperature limit for $k$ even it is sufficient to take the limit of $\tanh (\beta h_i)$ which reads:
\beq
\lim_{T\rightarrow 0} \tanh \beta h_i =\left\{
\begin{array}{ccc}
\sign k_i & {\rm for} & k_i \neq 0
\\
\tanh \e_i & {\rm for} & k_i=0
\end{array}
\right.
\eeq
The free energy shift reads \cite{MP2}
\beq
- \D F^{(1)}=|k_0|+\sum_{i=1}^{k+1}(|J_i|-|\sign k_i| )\ .
\eeq
We have determined the distribution of the fields by evolving numerically populations of triplets
$\{k_i,e_i,z_i\}$ at $u=0$.
For generic even values of $k$ the resulting distribution of the fields is such that: i) the $z$'s
have a finite probability of being exactly zero, ii) The non-zero values of $z$ are rather large and
diverging with the connectivity $k$, iii) there is a strong correlation between $z_i$ and $h_i$ such
that $\sign (h_i z_i)$ has a very small probability of being negative, iiii) the entropic
corrections $\e_i$ are rather large, (see fig.  \ref{figure1}).

\begin{figure}[htb]
\begin{center}
\epsfig{file=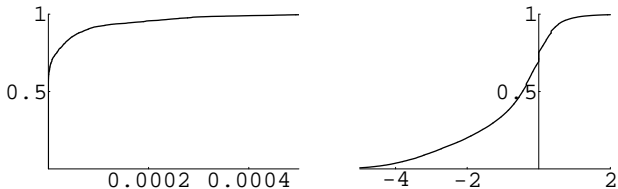,width=9cm} 
\caption{$T\rightarrow 0$ limit annealed theory for Bimodal Distribution, k=3. Left: cumulative distribution function of $\sign (k_i) z_i$ with rescaled $z_i$; Right: cumulative distribution function of the $\e_i$ at $k_i=1$}
\label{figure2}
\end{center}\end{figure}

{\it Odd Values of $k$} - In this case the fields generated in the merging of $k$ branches are
concentrated over odd values of the integers.  In particular they cannot be equal to zero and the
term $\tanh \beta h_i$ will be always equal to $\pm 1$.  This has deep consequences on the
reweighing term, eq.  (\ref{dx11}), and leads to the divergence of the $z$'fields.  For $k$'s
different from zero eq.  (\ref{dx11}) can be rewritten as the sum over $i$  terms of the form:

\beqd \Delta X_i=\left( \sign(h_0)(1-2 e^{-2\beta |h_0|})
-\sign(J_i)\sign(h_i) \right.  \times \eeqd \beq \times \left.  (1-2 e^{-2 \beta |J_i|}-2 e^{-2
\beta |h_i|})\right)\left| {du \over dh_i}\right| \sign(J_i)z_i
\label{cruc}
\eeq This term is zero if $z_i=0$, while if $z_i\neq 0$ it can be finite, null or equal to
$-\infty$, thus leading to a vanishing weight.  In order to analyze these possibilities it is useful
to distinguish two cases.  In the first case we have: \beq \sign(h_0)\sign(J_i)\sign(h_i)=-1 \eeq
Then the first factor in (\ref{cruc}) is finite and equal to $-2\sign(J_i)\sign(h_i)$ therefore \beq
\Delta X_i=- 2 \sign(h_i)\left| {du \over dh_i}\right|z_i \eeq Now we make the crucial hypotheses
that $z_i$ either is zero (with finite probability) or infinite (more precisely $O(e^{2 \beta
|J|})$) and in this case it has the same sign as $h_i$.  These hypotheses are motivated by the
behavior of the various quantities at finite low temperature observed numerically.  At the end we
will check that these hypotheses are self-consistent.  From eq.  (\ref{dudq}) we find that:
\beqd \Delta X_i=\left\{
\begin{array}{lcl}
O(e^{-2 \beta (|h_i|-|J_i|-1)})
\simeq -\infty & {\rm for} & |k_i|=1 , z\neq 0 
\\
O(e^{-2 \beta (|h_i|-|J_i|-1)})
\simeq 0  & {\rm for} & |k_i|>1, z\neq 0 
\end{array}
\right.
\eeqd
The last equation follows from the fact that the smaller value of $|k_i|>1$ is $|k_i|=3$
because of the odd value of $k$.
In the second case we have:
\beqd
\sign(h_0)\sign(J_i)\sign(h_i)=1
\eeqd
Then the first factor in (\ref{cruc}) is infinitesimal and $\Delta X_i$ is given by:
\beq
\Delta X_i=2\left( e^{-2 \beta |J_i|}- e^{-2\beta |h_0|} + e^{-2 \beta |h_i|})\right)\left| {du \over dh_i}\right| \sign(h_i)z_i
\label{cruc2}
\eeq
We note that the first term is order $e^{-2 \beta |J_i|}$ therefore 
\beqd
\Delta X_i=
O(e^{-2 \beta (|h_i|-1)})
\simeq 0  \ \ \ {\rm for} \  z\neq 0 ,|k_i|>1 
\eeqd
Finally we consider the case $|k_i|=1$. This is the only case in which we have a finite reweighing. 
To rewrite eq. (\ref{cruc2}) we note that the sign of $h_i$ is simply given by the sign of $k_i$ thus we have:
\beqd
|h_i|=|k_i|+T \sign(k_i)\e_i \Longrightarrow  e^{-2 \beta |h_i|}=e^{-2 \beta |k_i|}e^{-2 \sign(k_i)\e_i}
\eeqd
The same result applies also to $h_0$ thus we have:
\beqd
\Delta X_i=2\left| {du \over dh_i}\right| \sign(h_i)z_i e^{-2 \beta |J_i|} \ \times
\eeqd
\beqd
 \times \left\{
 \begin{array}{lcl}
  1+ e^{-2 \sign(k_i)\e_i}
&{\rm for} &  |k_i|=1  , |k_0|>1 
\\
  1-e^{-2 \sign(k_0)\e_0}+ e^{-2 \sign(k_i)\e_i}
& {\rm for}  & |k_i|=1  , |k_0|=1 
\end{array}
\right.  \eeqd The presence of the term $O(e^{-2 \beta |J_i|})$ fixes the scales of the non-zero
values of $z$ to $O(e^{2 \beta |J_i|})$ in agreement with our hypotheses.  Therefore we must rescale
$z$ to work directly in $T\rightarrow 0$ limit.  The condition $\sign(h_iz_i) \geq 0 $ is
self-consistent under iteration, indeed it can be checked that the reweighing $\Delta X_{i}$ gives zero
weight to the triplet $\{k_0,\e_0,z_0\}$ unless each contribution $du_i/dh_i z_i$ to $z_0$ has the
same sign as $h_0$.  In order to compute the complexity we must consider also the merging of $k+1$
branches.  The main difference is that in this case $k_0$ can be equal to zero.  First of all we
note that if $k_0\neq 0$ the computation of the anomalous shift $\Delta X$ is the same as above.
The quantity $\Delta X_{i}$ in the case that $k_0=0$ can be rewritten as the sum over $i$ of the
following terms in the low temperature limit: 
\beqa \Delta X_i & = &\left( \tanh \e_0
-\sign(J_i)\sign(h_i)\right)\left| {du \over dh_i}\right| \sign(J_i)z_i= \nonumber \\
&=& -\left| {du \over dh_i}\right| \sign(h_i)z_i(1-\sign(J_i h_i)\tanh \e_0 )
\eeqa
 since $|\tanh \e_0|\leq 1$ the last factor is always positive and
\beq
\Delta X_i=
-\infty \ \ \  {\rm for} \   z_i\neq 0, k_0=0  
\eeq
The same result applies when considering the quantities which enter the variational
expression of the complexity \cite{R2} and of the average free energy.
We have determined the distribution of the fields at $u=0$ for odd $k$ by evolving numerically
populations of triplets $\{k_i,e_i,z_i\}$, where $z_i$ has been rescaled as $z_i e^{-2\beta
|J|}\rightarrow z_i$ in order to be finite, see fig.  (\ref{figure2}).  As expected there is a
finite probability of having $z_i=0$.
We stress that the complicated behavior of the odd-$k$ model is determined by the fact that $k_i$ is
always different from zero, therefore more generic situation ({\it e.g.} fluctuating connectivity)
should display a simpler behavior, like the one of the even-$k$ model.

In \cite{R2} it was found that in the Bethe lattice spin-glass the total complexity (which is
obtained setting $u=0$) is given by the SUSY-breaking theory {\it i.e.} $z\neq 0$.  We have checked
that this property remains true in the $T\rightarrow 0$ limit in the three models considered.  We
also stress that if we force $z_i=0 $ the population dynamics algorithms of the bimodal
case do not converge and in particular the $\epsilon_i$'s diverge implying that the decomposition
$h_i=k_i+T\epsilon_i$ breaks down leading to a distribution which is no more concentrated over the
integers.  The numerical computation of the complexity requires higher computational efforts and is
currently underway.

Our results raise an interesting question.  In the bimodal case the ``strong" variables $k_i$ obey
the recursive equations eq.  (\ref{reck}) that do not depend on the $e_i$'s and the $z_i$'s.  We
call the solutions $\{k_i^{\a}\}$ of these equations ``strong belief-propagation solutions" (SBPS).
Instead the complexity we are computing yields the number of solutions
$\{k_i^{\a},e_i^{\a},z_i^{\a}\}$ of equations (\ref{reck},\ref{rece},\ref{recz}), we call them full
belief-propagation solutions (FBPS).  In the standard theory with $z_i=0$ the $X$ shift is zero and
the complexity depends only on the strong variables, thus we obtain for the FBPS complexity the same
result we would obtain for the complexity of the SBPS. Instead the fact that when $z_i\neq 0$ the
complexity depends not only on the $k_i$ but also on the $\e_i$ and $z_i$ leads to the possibility
that there is no biunivocal correspondence between FBPS and SBPS. This could happens either because
for each SBPS there are many FBPS or because there is no FBPS corresponding to a given SBPS.

However it is interesting to notice that the question can be formulated as a percolation problem.
Starting from a given SBPS, $\{k_i^{\a}\}$, we want to determine a FBPS $\{\e_i^{\a},z_i^{\a}\}$
which solves equations (\ref{reck},\ref{rece},\ref{recz}).  We can proceed in the following way, we
consider a root site $0$ and we start exploring all the trees starting from it.  On each spin $i$ we
measure the field $k_i$ that it passes in the direction of the root.  If $|k_i|\leq 1$ we go to the
next level of the tree while if $|k_i|>1$ we stops because from eqs.  (\ref{ge}) and (\ref{dudq}) we
knows that it will pass zero messages $g(k_i,\e_i)=0$ and $du_i/dh_i z_i=0$.  If at a certain point
we stop completely, we are left with a tree with vanishing messages passed by the leaves, thus we
can compute the unique value $(\e_0,z_0=0)$ in a sweep.  If we can repeat the process for each point
of the lattice we find that there is a unique FBPS corresponding to the given SBPS and that it has
$z_i=0\, \forall i$.  Thus we can find non vanishing $z$-fields only if we have percolation in this
process, correspondingly we cannot determine uniquely $\e_0$ and the biunivocal correspondence
between SBPS and FBPS becomes uncertain.

We thank A. Montanari for interesting discussions.Work supported in part by the European Community's Human
Potential Program under the project Dyglagemem and Evergrow.


\begin{thebibliography}{99}

\bibitem{MPV}        M.~M\'ezard, G.~Parisi, and M.A.~Virasoro,
		  ``Spin glass theory and beyond", World Scientific (Singapore 1987).


\bibitem{isp}
				 A. Crisanti, L. Leuzzi, and T. Rizzo, {\sf cond-mat/ 0406649}, A. Annibale, G. Gualdi, A. Cavagna {\sf cond-mat/ 0406466}. 

\bibitem{noiquen}
A. Crisanti, L. Leuzzi, G. Parisi, T. Rizzo, 
Phys. Rev. Lett. {\bf  92}, 127203 (2004)


\bibitem{BMan}A.J. Bray and  M.A. Moore, J. Phys. C {\bf 13}, L469 (1980).


\bibitem{var} J. Kurchan,. J. Phys. A {\bf 24} (1991) 4969, A. Cavagna, I. Giardina, G. Parisi and M. M\'ezard, J. Phys. A 
{\bf 36} (2003) 1175, A. Crisanti, L. Leuzzi, G. Parisi, T. Rizzo, Phys. Rev. B, 68 (2003) 174401


\bibitem{ABM} T. Aspelmeier, A. J. Bray, and M. A. Moore
Phys. Rev. Lett. {\bf 92}, 087203 (2004).

\bibitem{PR} 
			  G. Parisi and T. Rizzo, J. Phys. A 37 (2004) 7979

\bibitem{CGPprl}A. Cavagna, I. Giardina, G. Parisi,
					Phys. Rev. Lett. 92, 120603 (2004)
 
\bibitem{R1}
			  T. Rizzo  {\sf cond-mat/0403261}.


\bibitem{MP1}           M.~M\'ezard and G.~Parisi, 
		  Eur. Phys. J. B {\bf 20} , 217 (2001)

\bibitem{MP2}   M.~M\'ezard and G.~Parisi, 
			J. Stat. Phys 111 (2003) 1

\bibitem{MPZ}
			M. M\'ezard, G. Parisi and R. Zecchina, {\it Science}  297:812, 2002


\bibitem{Moncomp}
			R. Monasson, Phys. Rev. Lett. {\bf 75} (1995) 2847


\bibitem{MRpMRP} A. Montanari and F. Ricci-Tersenghi,
		  Eur. Phys. J. B 33, 339 (2003). A. Montanari, G. Parisi and F. Ricci-Tersenghi, J. Phys. A 37, 2073 (2004).

\bibitem{CGPcav} 
				 A. Cavagna, I. Giardina, G. Parisi,
				  {\sf cond-mat/0407440} 


\bibitem{R2}    T. Rizzo  {\sf cond-mat/0404729} 



\end{thebibliography}
\end{document}